\documentclass[aps,prl,twocolumn,showpacs,showkeys,footinbib,superscriptaddress]{revtex4-2}
\usepackage{fontspec}

\usepackage{amsmath}
\usepackage{amssymb}
\usepackage{graphicx}
\usepackage{bm}
\usepackage{color}
\usepackage{relsize}
\usepackage{braket}
\usepackage[caption=false]{subfig}
\usepackage{hyperref}
\usepackage[all]{hypcap}
\usepackage{siunitx}

\renewcommand{\vec}[1]{\mathbf{#1}}

\begin{document}
\title{Magnetically induced Circular Photogalvanic Effect in Symmetric Two-dimensional Materials}
\date{\today} 

\author{Peng Liu}
\affiliation{Department of Physics, Hangzhou Dianzi University, Hangzhou, Zhejiang 310018, China}
\author{Fanhao Jia}
\affiliation{Department of Physics, Hangzhou Dianzi University, Hangzhou, Zhejiang 310018, China}
\author{Ruixue Li}
\affiliation{Department of Physics, Hangzhou Dianzi University, Hangzhou, Zhejiang 310018, China}
\author{Yuan Li}
\affiliation{Department of Physics, Hangzhou Dianzi University, Hangzhou, Zhejiang 310018, China}
\author{Igor \v{Z}uti\'c}
\email{zigor@buffalo.edu}
\affiliation{Department of Physics, University at Buffalo, State University of New York, Buffalo, NY 14260, USA}
\author{Gaofeng Xu}
\email{xug@hdu.edu.cn}
\affiliation{Department of Physics, Hangzhou Dianzi University, Hangzhou, Zhejiang 310018, China}

\begin{abstract}
Photocurrents that depend on the helicity of the incident light can be generated in both bulk and low-dimensional materials lacking inversion symmetry, known as the circular photogalvanic effect (CPGE). We propose that by employing a magnetic effect, the limitation on the inversion symmetry broken materials can be overcome, such that helicity-dependent photocurrent can be generated in a symmetric material, i.e., a magneto-circular photogalvanic effect (MCPGE). As a proof of principle, we elucidate the mechanism of such an MCPGE through an effective Hamiltonian of a monolayer SbH on a magnetic substrate with an adjustable magnetization. Moreover, the associated response in optical absorption is analyzed, both single-particle and excitonic, through a Bethe-Salpeter equation to describe the Coulomb interaction in excitons. Our result broadens the mechanism of CPGE and opens new opportunities for optoelectronic devices. 
\end{abstract}
\maketitle



Unlike photovoltaic effect in \emph{p-n} junctions, the circular photogalvanic effect (CPGE) occurs in homogeneous materials without inversion symmetry, where photocurrents are dependent on the circular polarization of the incident light~\cite{Ivchenko1978:JETP, Belinicher1978:PLA, Belinicher1980, Ganichev2001:PRL, Ganichev2003:JP}. 
As an actively studied topic in nonlinear optoelectronic responses, CPGE has attracted numerous research attention for both fundamental physics and its advantages in spintronics and optoelectronics~\cite{Zutic2004, Zhou2018:npj}. CPGE has been observed in a variety of materials, including bulk and two-dimensional (2D) materials, such as quantum wells~\cite{Belkov2005,Belkov2008}, Weyl semimetals~\cite{deJuan2017:NC,Ma2019:NM, BaiWeyl2025,Chen2024Weyl}, monolayer (ML) transition metal dichalcogenides~\cite{Zhang2014TMD, Sun2023TMD, Xu2021:npj}, and topological insulators~\cite{McIver2012:NN, Okada2016:PRB}. 


Generally, inversion symmetry breaking of the lattice structure, either intrinsic or extrinsic induced by gradients of strain~\cite{Jiang2021:NN, Ji2025:AM} and defects~\cite{Jo2022:NC}, is essential to support CPGE, which arises from the asymmetric optical excitation of photocurrents. As an exception, CPGE was observed in a symmetric material, a silicon nanowire, where the inversion symmetry is locally broken by the electric field at the metal contact~\cite{Dhara2015}. The electric field leads to a splitting of the valence bands, which leads to asymmetric optical excitation. 


Inspired by the electric-field effect, we propose that magnetic effects can be exploited to induce CPGE in centrosymmetric materials, by breaking the symmetry in electronic band structures, rather than changing the lattice structures deliberately. Specifically, we illustrate such a magneto-circular photogalvanic effect (MCPGE) and associated optical absorption as well as their tunability by the strength and direction of an effective magnetic field in a centrosymmetric 2D material. 

\begin{figure}[h]
\includegraphics[scale =0.39]{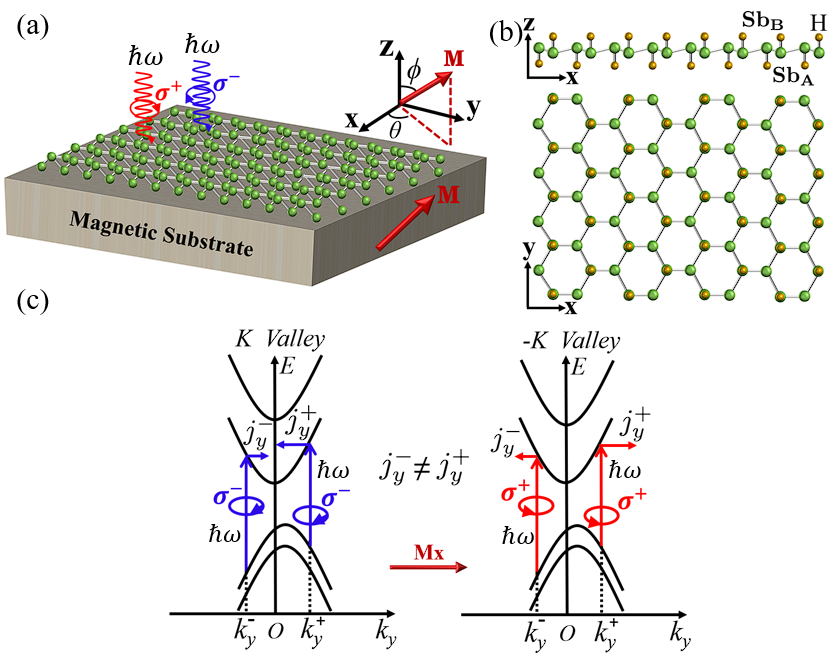}
\caption{(a) Schematic of an atomic ML on a magnetic substrate with a tunable direction and strength of magnetization illuminated by circularly polarized light. (b) Top and side views of a ML SbH. (c) Band structure and valley-contrast optical transitions at the $K$ and $-K$ valleys under a magnetic exchange field along the $x$ direction. The inversion symmetry of the energy dispersion is broken, which leads to unbalanced photocurrents ($j_y^+ \ne j_y^-$) and thus a nonzero photocurrent.
\label{fig1}}
\end{figure}





To exemplify the MCPGE, we consider a ML SbH on a magnetic substrate with a tunable magnetization, as shown in Fig.~\ref{fig1}(a). For a 2D material, with its thickness fully covered by penetration lengths of proximity effects, a magnetic proximity effect can serve as an efficient approach to introduce magnetism in the entirety of the material~\cite{Hauser1969, Scharf2017:PRL, Zutic2018proximity, Choi2023:NM, Zhou2023proximity, Wang2024:PRB}.  A ML SbH has been proposed to be a large-gap topological insulator, with a hexagonal lattice structure as shown in Fig.~\ref{fig1}(b)~\cite{Song2014:NPGAM}. Due to the buckled structure, Sb atoms can be divided into $A$ and $B$ sublattices [Fig.~\ref{fig1}(b)] with a vertical separation (buckling height)~\cite{Zhou2016:PRB}. 
Consequently, staggered magnetic exchange fields are induced in $A$ and $B$ sublattices by the substrate, such that the symmetry between the two sublattices are broken~\cite{Xu2020:PRL}. Notably, such an effective inversion symmetry breaking of the ML SbH does not require an explicit change of its lattice structure. In addition, it also leads to inequivalent $K$ and $-K$ valleys at the corners of the hexagonal Brillouin zone with valley-contrast Berry curvature and circular dichroism: $K$ ($-K$) valley is exclusively coupled to $\sigma^+$ ($\sigma^-$) helicity of light~\cite{Yao2008Va,Cao2012Va,Sharma2024Va,Yang2024:PRB}.

A ML SbH has direct band gaps at $\pm K$ valleys, as depicted in Fig.~\ref{fig1}(c). With an in-plane magnetic exchange field from the substrate, the originally symmetric energy bands at the valleys are shifted along the direction perpendicular to the field. The energy dispersion is influenced by in-plane magnetic fields in such a non-trivial way due to a competition between the Zeeman spin splitting and Rashba spin-orbit coupling (SOC)~\cite{Wu2017:JAP, Akbari2022:PRR,Note:SM}. Consequently, optical transitions at each valley become asymmetric with respect to the corresponding $K$ ($-K$) point: $|k_y^-| \neq |k_y^+|$, schematically shown in Fig.~ \ref{fig1}(c). Such an asymmetry prevents the cancellation of the photocurrent and leads to a net current due to unbalanced currents ($|j_y^+| \neq |j_y^-|$) of electrons excited at $k_y^{\pm}$ by a circularly polarized light of a certain frequency. Since $\pm K$ valleys have opposite Berry curvatures, the net currents at the two valleys are in opposite directions. Combined with the valley-contrast circular dichroism, a particular helicity of light leads to excitation of carriers in the exclusively coupled valley and a photocurrent in the valley-contrast direction, i.e., the MCPGE.

Note that the MCPGE proposed here is fundamentally different from the following two effects in the mechanisms. In magneto-gyrotropic effects, the generation of photocurrents requires simultaneously a magnetic field and gyrotropic media, which lacks inversion symmetry~\cite{Belkov2005, Belkov2008}. A magnetic photogalvanic effect is proposed in a magnetic material CrI$_3$, where the magnetism comes from the magnetic material itself, and photocurrent is forced to vanish under circularly polarized light and only exists under linearly polarized light~\cite{Zhang2019:NC}. 



The band gaps of a ML SbH on a magnetic substrate locate at the $\pm K$ valleys [Fig.~\ref{fig1}(a)], for which a low-energy effective Hamiltonian can be constructed. The orbital basis functions at the $A$ and $B$ sublattices is given by $|\Phi_{A, \tau} \rangle=|- i \tau p_x^A  + p_y^A \rangle $ and
$|\Phi_{B, \tau} \rangle =| i \tau p_x^B + p_y^B \rangle$, respectively, with $\tau = \pm 1$ indices for $\pm K$ valleys. Therefore, a spinful Hilbert space $\{ | \Phi_{A,\tau}^{\uparrow}\rangle,  |\Phi_{B,\tau}^{\uparrow}\rangle,  |\Phi_{A,\tau}^{\downarrow}\rangle, |\Phi_{B,\tau}^{\downarrow}  \rangle\}$ is constructed. The total Hamiltonian $H_\mathrm{tot}$ can be expressed as $H_\mathrm{tot} = H_0 + H_\mathrm{ex} + H_U + H_R$, consisting of terms from the bare ML, magnetic exchange, staggered potential, and Rashba spin-orbit coupling (SOC)~\cite{Dominguez2018:PRB, Xu2020:PRL}. Specifically, $H_0$ describing the bare ML is ~\cite{Zhou2015:NL,Zhou2016:PRB,Zhou2018:npj,Song2014:NPGAM}
\begin{eqnarray}
H_0 = \hbar v_F (k_x \sigma_x \tau_z + k_y \sigma_y) + \lambda_{SO} \sigma_z S_z \tau_z, 
\label{1.1}
\end{eqnarray}
where $\sigma_i$, $\tau_z$ and $S_z$ are Pauli matrices for the orbital, valley, and spin degrees of freedom, respectively. $v_F$ is the Fermi velocity, $\lambda_{SO}$ denotes the effective on-site SOC, and $k_{x,y}$ are momenta measured from $\pm K$ points. The magnetic proximity effect induced by the magnetic substrate leads to staggered exchange fields $M_{A,B}$ for the $A$ and $B$ sublattices separately due to their vertical distance~\cite{Choi2023:NM, Zhou2023proximity}, such that 
\begin{eqnarray}
H_\mathrm{ex} = \hat{\mathbf{n}} \cdot \mathbf{S}[M_A (1 + \sigma_z)/2 + M_B (1 -\sigma_z)/2], 
\label{1.2}
\end{eqnarray}
where the unit vector $\hat{\mathbf{n}}$ indicates the direction of the substrate magnetization ($\vec M =M \hat{\mathbf{n}}$) and $\mathbf{S}$ is the vector of spin Pauli matrices. Besides, the existence of the substrate inevitably induces a relatively small but nonzero staggered potential $U$, such that $H_U = U \sigma_z$, even without external electric fields~\cite{Xu2020:PRL}. The Rashba SOC is $H_R = 3 \lambda_R (\sigma_x S_y \tau_z - \sigma_y S_x)$, with $\lambda_R$ the Rashba SOC parameter~\cite{Dominguez2018:PRB}. 

An injection current can be induced by a circularly polarized light through a second-order response, with a generation rate~\cite{Sipe2000:PRB, deJuan2017:NC, Xu2021:NC, Yang2024:PRB}
\begin{eqnarray}
d j_{i}/dt=\beta_{i j}(\omega)[ \vec E(\omega)\times \vec E^{*}(\omega)]_j, 
\label{2}
\end{eqnarray}
where $\vec E(\omega)=\vec E^*(-\omega)$ is the optical electric field, $\beta_{i j}$ is the CPGE tensor, with $i$ and $j$ denoting cartesian indices $x$, $y$ and $z$. After a saturation time $\tau_s$, the injection current reaches a steady-state value $j_i =\tau_s \beta_{i j}(\omega)[ \vec E(\omega)\times \vec E^{*}(\omega)]_j$. We consider a relatively small frequency of the incident light, such that interband optical transition only exists between the lower conduction band (CB) and higher valence band (VB). Consequently, a two-band model involving the lower CB and higher VB can be applied to describe the CPGE, such that the CPGE tensor for a 2D material can be given by~\cite{Sipe2000:PRB, deJuan2017:NC, Yang2024:PRB}
\begin{eqnarray}
\beta_{i j}(\omega)=\sum_{\mathbf{k} } \frac{i \pi e^{3}}{\hbar^{2} A} \partial_{k_{i}} E_{\mathbf{k}, 12} \Omega^{v}_j(\mathbf{k}) \delta\left(\hbar \omega-E_{\mathbf{k}, 21}\right), 
\label{3}
\label{Eq:CPGE}
\end{eqnarray}
where $A$ is the area of the 2D sample, $E_{\mathbf{k},mn}=E_{\mathbf{k},m}-E_{\mathbf{k},n}$ is the energy difference of the two bands, with $m, n=1, 2$ corresponding to VB and CB, and $\Omega^{v}_j(\mathbf{k})$ is the Berry curvature of the VB~\cite{Xiao2010:RMP}. The $\delta$ function determines the contour in the 2D $\mathbf{k}$ space where resonant transition occurs, which is modeled as a Lorentzian with broadening $\Gamma$ in numerical calculations~\cite{Note:SM}.



\begin{figure*}[t]
\includegraphics[scale=0.32]{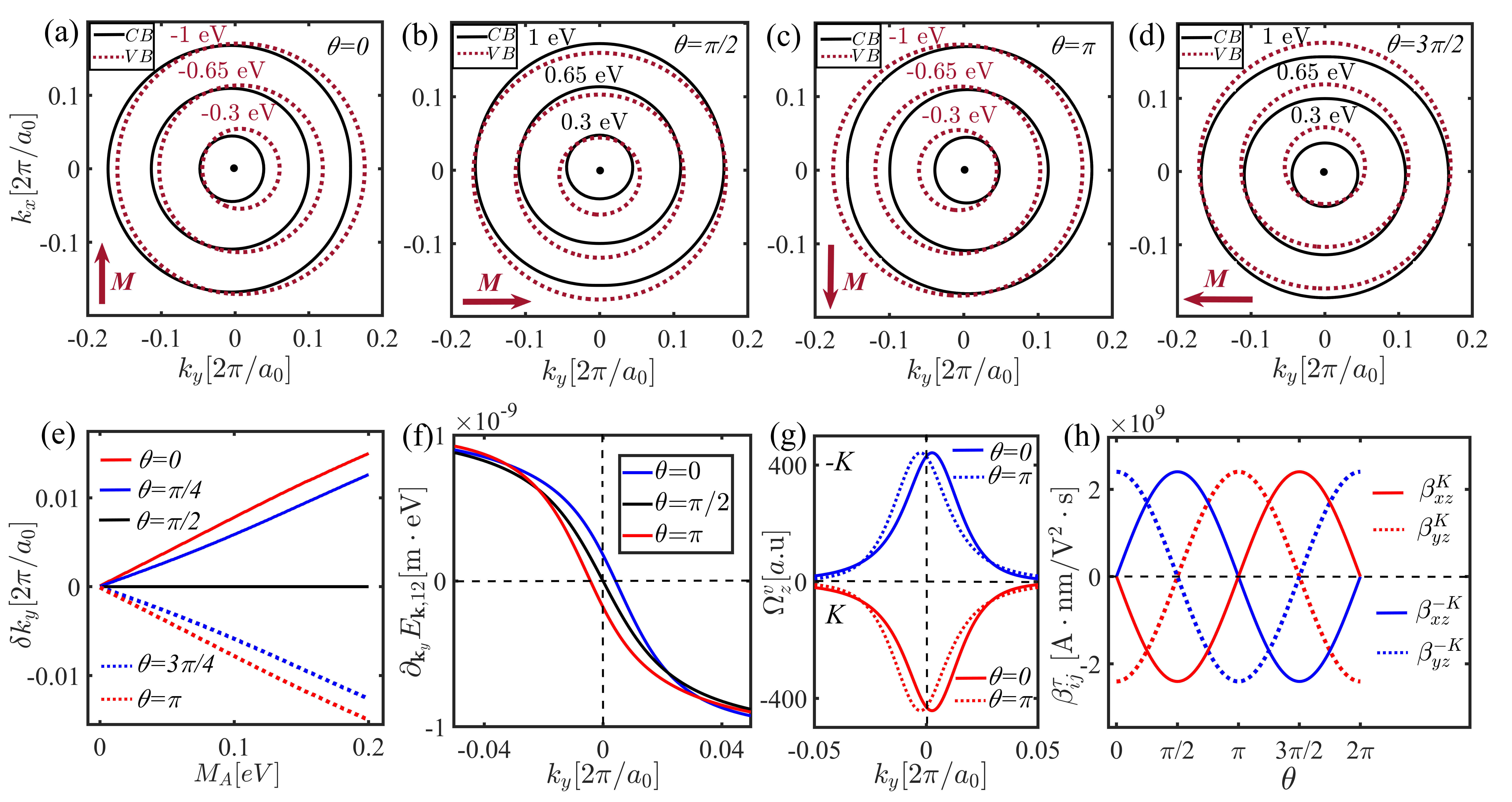}
\caption{\label{fig2}(a)-(d) Isoenergy contours showing shifts of conduction (solid) and valence bands (dashed) at the $\pm K$ valleys under in-plane magnetic exchange fields along various directions ($\theta=0, \pi/2, \pi, 3\pi/2$), depicted by the arrows. (e) Relative shifts between the $k_y$ values of the valence band maximum (VBM) and conduction band minimum (CBM), $\delta k_y=k_y$(VBM)$-$$k_y$(CBM), as functions of $M_A$ along different directions ($\theta=0, \pi/4, \pi/2, 3\pi/4, \pi$). 
(f) $\partial_{k_{y}} E_{\mathbf{k}, 12}$ as functions of $k_y$ for different directions of $\vec M$ ($\theta=0, \pi/2, \pi$). 
(g) Berry curvatures $\Omega_z^v$ of the valence band at the $\pm K$ valleys for $\vec M$ along $\theta=0$ and $\pi$. (h) Evolution of CPGE tensors $\beta_{xz}^{\tau}$ (solid) and $\beta_{yz}^{\tau}$ (dashed) at the $K$  (red) and $-K$ valleys (blue) with $\theta$.  }
\vspace{0.7cm}
\end{figure*}


Moreover, we investigate the associated optical absorption, both single-particle and excitonic. In particular, the optical absorption of 2D materials is dominated by excitons due to 2D confinement and reduced dielectric screening~\cite{Wang2018:RMP}. An exciton state can be defined as $\ket{\Psi^{\tau}_{S}}=\sum_{vc\vec{k}}\mathcal{A}^{S\tau}_{vc\vec{k}}\hat{c}^\dagger_{\tau c\vec{k}}\hat{c}_{\tau v\vec{k}}\ket{\mathrm{GS}}$ with $\mathcal{A}^{S\tau}_{vc\vec{k}}$ the exciton envelope function, $\hat{c}^\dagger_{\tau c\vec{k}}$ ($\hat{c}_{\tau v\vec{k}}$) the creation (annihilation) operator of an electron in a CB $c$ (VB $v$) in the valley $\tau$, and $\ket{\mathrm{GS}}$ the ground state with fully occupied VBs and unoccupied CBs. The influence of magnetic exchange fields on excitons can be analyzed by numerically solving the Bethe-Salpeter equation (BSE)~\cite{Rohlfing2000:PRB, Scharf2016:PRB, Scharf2017:PRL, Cao2024:PRB, Cao2025:PRL}
\begin{equation}
\left[\Omega^{S \tau}-\epsilon_{c}^{\tau}(\boldsymbol{k})+\epsilon_{v}^{\tau}(\boldsymbol{k})\right] \mathcal{A}_{v c k}^{S \tau}=\sum_{v^{\prime} c^{\prime} k^{\prime}} \mathcal{K}_{v c k, v^{\prime} c^{\prime} k^{\prime}}^{\tau} \mathcal{A}_{v^{\prime} c^{\prime} \boldsymbol{k}^{\prime}}^{S \tau}, 
\label{Eq:BSE}
\end{equation}
where $\Omega^{S \tau}$ is the energy of the exciton state, the energies $\epsilon_n^\tau$ ($n=c ,v$) and their corresponding eigenstates $\eta_n^\tau$ are obtained from the single-particle Hamiltonian, $H_\mathrm{tot}^\tau\eta_{n\boldsymbol{k}}^\tau=\epsilon_{n}^{\tau}(\boldsymbol{k})\eta_{n\boldsymbol{k}}^\tau$. The interaction kernel $\mathcal{K}^{\tau}_{vc\vec{k},v'c'\vec{k}'}$ incorporates the many-body Coulomb interaction between electrons in the ML, determined by the dielectric environment~\cite{Keldysh1979:JETP,Cudazzo2011:PRB,Scharf2019:JPCM,Note:SM}. With the exciton envelope functions $\mathcal{A}_{v c k}^{S \tau}$, the absorption spectra can be further achieved
\begin{align}
\alpha^{ \pm}(\omega)=\frac{4 e^{2} \pi^{2}}{c \omega} \frac{1}{A} \sum_{S \tau}\left|\sum_{v c k} \mathcal{D}_{v c k}^{\tau, \pm} \mathcal{A}_{v c k}^{S \tau}\right|^{2} \delta\left(\hbar \omega-\Omega^{S \tau}\right)
\label{5}
\end{align}
where $\omega$ denotes the frequency of light propagating along the \(-z\) direction, and \( c \) is the speed of light. The velocity matrix element for left/right circularly polarized light is given by  
$\mathcal{D}_{v c k}^{\tau, \pm} = [\eta_{v \mathbf{k}}^{\tau}]^{\dagger} \hat{v}_\pm \eta_{c \mathbf{k}}^{\tau}$,
where $\hat{v}_\pm=(\hat{v}_x \pm i\hat{v}_y)/\sqrt{2}$ and $\hat{v}_{x/y}=\partial H_{tot}/\partial(\hbar k_{x/y})$. The $\delta$ function is modeled by a Lorentzian with broadening $\Gamma$.




To illustrate the effect of the in-plane magnetic exchange field on the energy bands, we show in Fig.~\ref{fig2}(a-d) the isoenergy contours of CBs and VBs at $\pm K$ valleys with the in-plane magnetic exchange field along different directions, depicted by the arrows. The shifts of VBs are more significant than those of the CBs,  which is a consequence of $M_A \gg M_B$. 
Therefore, the CBs are nearly unshifted, and we can focus on the relative shifts of VBs, which are always perpendicular to the direction of the in-plane magnetic exchange field, at a direction angle $\theta'=\theta + \pi/2$. As we have examined, such a non-trivial influence on energy dispersion by in-plane magnetic fields is a result of the competition between the Zeeman spin splitting and Rashba SOC in the system. 
Due to the degeneracy of energy bands in $\pm K$ valleys under in-plane fields, isoenergy contours are identical in the two valleys. 
Notably, since the direction of band shifts are the same with respect to the $\pm K$ points in the two valleys [see Fig.~\ref{fig1}(c)], the $\mathbf{k}$-space inversion symmetry of the energy bands from $+\mathbf{q}$ to $-\mathbf{q}$ are actually broken, with $\mathbf{q}$ the momenta measured from the $\Gamma$ point ($\mathbf{q}=\mathbf{k}\pm K$). 

To show the band shifts more explicitly, we present in Fig.~\ref{fig2}(e) the relative shift between the $k_y$ values of the valence band maximum (VBM) and conduction band minimum (CBM), $\delta k_y=k_y$(VBM) $-$ $k_y$(CBM), with increasing $M_A$, along different directions $\theta=0, \pi/4, \pi/2, 3\pi/4, \pi$. 
Generally, the magnitudes of the shifts exhibit monotonic increases with larger $M_A$, except for $\theta=\pi/2$, for which the shift vanishes since $\vec M$ is perpendicular to $x$ direction. The direction of the shifts undergoes a reversal as the $\theta$ goes beyond $\pi/2$. Therefore, the magnitude and direction of the shifts are determined by the projection of $\vec M$ along $x$ direction. Note that a constant ratio $M_A/M_B$ has been kept when increasing $M_A$. 

To elucidate the effects of a magnetic exchange field on CPGE tensors, we further examine numerically the influences on their components $\partial_{k_{y}} E_{\mathbf{k}, 12}$ and Berry curvature $\Omega^{v}_j(\mathbf{k})$, which are consequences of the shift of energy bands.
As shown in Fig.~\ref{fig2}(f), for $\vec M$ along $y$ direction ($\theta= \pi/2$), $\partial_{k_{y}} E_{\mathbf{k}, 12}$ is an odd function of $k_y$, since energy bands remain symmetric along $k_y$ [Fig.~\ref{fig2}(b)]. 
However, when $\vec M$ is applied along the $x$ direction ($\theta= 0$), the curve is shifted towards $k_y$ and becomes asymmetric, due to the shift of energy bands [Fig.~\ref{fig2}(a)]. The curve for $\theta = \pi$ is shifted towards the opposite direction, showing the dependence of the shift direction on the $x$ component of $\vec M$. 
Similarly, Berry curvatures are shifted towards opposite directions for  $\theta =0$ and $\pi$ and become asymmetric with respect to the $\pm K$ points, as shown in Fig.~\ref{fig2}(g). Berry curvatures at the $\pm K$ valleys have a sign difference, i.e., $\Omega_{\vec k}^{K} = -\Omega_{\vec k}^{-K} $. The shifts $\delta k_y$, $\partial_{k_y} E_{\mathbf{k}, 12}$, and Berry curvatures are identical for $\pm K$ valleys. 

As a result, the asymmetric $\partial_{k_y} E_{\mathbf{k}, 12}$ and Berry curvature caused by the in-plane magnetic exchange field leads to nonzero CPGE tensors. As shown in Fig.~\ref{fig2}(h), $|\beta_{yz}^{\tau}|$ reaches maxima when $\theta = 0$, and gradually decreases to 0 as $\theta$ goes to $\pi/2$. The trend for $|\beta_{xz}^{\tau}|$ is reversed, which is connected to the shifts on the orthogonal direction. Besides, the CPGE tensors at $\pm K$ valleys are opposite, $\beta_{ij}^K=-\beta_{ij}^{-K}$, which is due to opposite signs of Berry curvatures. Since each valley is excited by its coupled helicity of light, the direction of injection currents generated by different helicities of light are opposite, leading to the CPGE. 

\begin{figure}[t]
\centering
\includegraphics[scale=0.32]{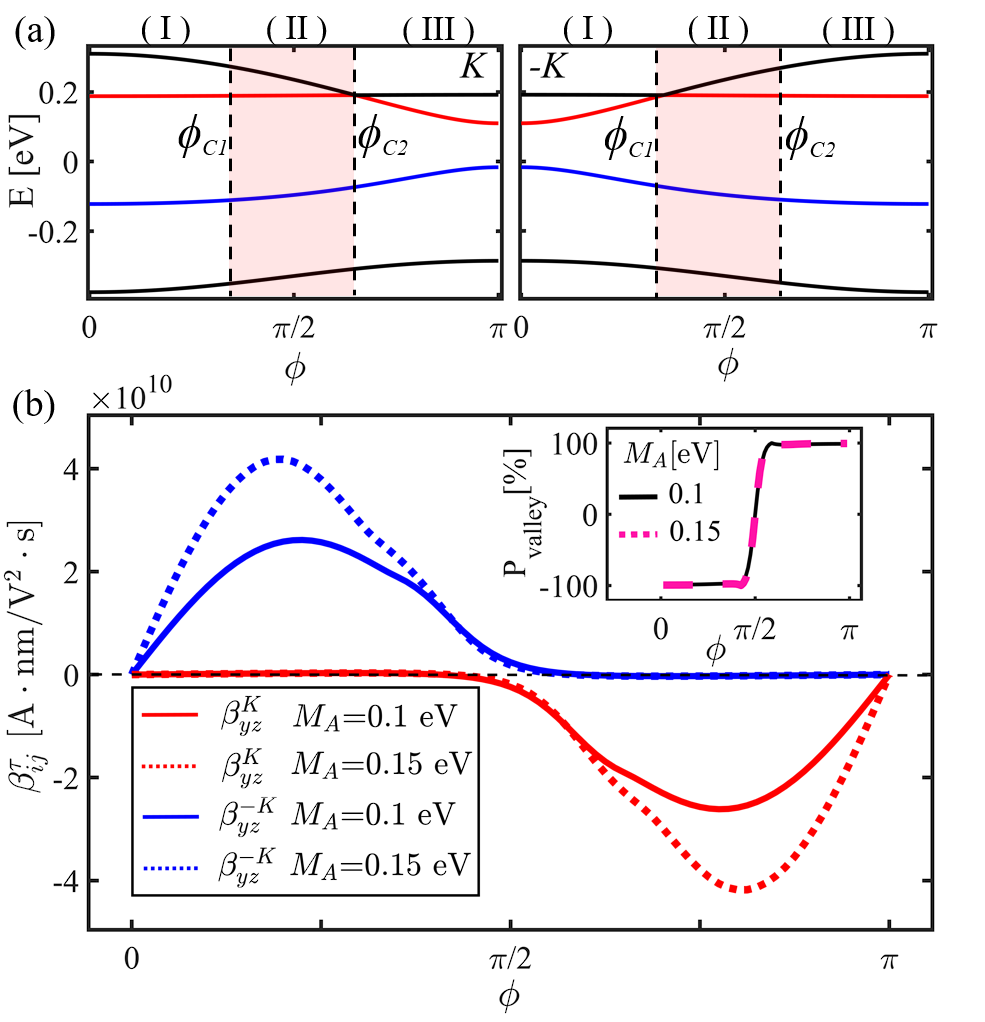}
\caption{(a) Evolution of band-edge energies with $\phi$ at the $\pm K$ valleys. (b) Evolution of CPGE tensors $\beta_{yz}^{\tau}$ at the $\pm K$ valleys with the direction of magnetic exchange field $\phi$ for different strengths of $M_A=0.1$ and $0.15$ eV. Inset: the valley polarization as a function of $\phi$. The photon energy $\hbar \omega=0.29$ eV, $\theta=0$. }
\label{fig3}
\end{figure}

We next investigate effects of exchange fields along arbitrary directions, which can be experimentally feasible through tilting substrate magnetization~\cite{Khang2018:NM, Hsu2024:NC, Hu2024:NPG}. 
An arbitrary direction of the exchange field gives rise to combined effects of its in-plane and out-of-plane components. 
The effect of the in-plane component on the CPGE tensor at one particular valley is essentially the same as analyzed above. However, the out-of-plane component breaks the valley degeneracy, as illustrated by the evolution of band-edge energies with $\phi$ at the $\pm K$ valleys in Fig.~\ref{fig3}(a). We see that the band gap at the $K$ valley decreases with increasing $\phi$, while that at the $-K$ valley increases. 
Consequently, the CPGE tensors at different valleys no longer have the same magnitudes and only differ by a sign as in Fig.~\ref{fig2}(h). In contrast, $\beta_{yz}^{\tau}$ can be strikingly different at the two valleys, as shown in Fig.~\ref{fig3}(b). Without loss of generality, a fixed energy of the incident photons is set as $\hbar \omega=0.29$ eV, which equals to the energy gap under in-plane exchange field ($\phi=\pi/2$); the in-plane component of the field is along the $x$ direction ($\theta=0$). For $0<\phi<\pi/2$, since $\hbar \omega$ is smaller than the energy gap in the $K$ valley, which forbids optical transitions, $\beta_{yz}^K$ is negligibly small. While $\hbar \omega$ is greater than the gap in the $-K$ valley, therefore $\beta_{yz}^{-K}$ is significantly larger than $\beta_{yz}^{K}$. The situation is reversed for $\pi/2<\phi<\pi$, where $\left |\beta_{yz}^{K}\right| \gg \left |\beta_{yz}^{-K}\right|$.

Such an asymmetry can be further illustrated by the valley polarization, defined as $P_{\text{valley}} =\left ( \left |\beta_{ij}^K \right| - \left |\beta_{ij}^{-K} \right| \right) / \left( \left |\beta_{ij}^K \right| + \left|\beta_{ij}^{-K} \right| \right)$, 
which is close to $\pm 100\%$ except for the transitional region near $\phi= \pi/2$, as shown in the inset of Fig.~\ref{fig3}(b). Due to this valley asymmetry, linearly polarized light is sufficient to induce a nonzero injection current, which can be further tuned by the magnitude and direction of the magnetic exchange field~\cite{Zhou2016:PRB}, as well as the photon energy through a resonance condition. 
Note that the CPGE tensors here are in comparable magnitudes to those in ML BiAsI$_2$, which stems from the broken inversion symmetry of the material~\cite{Yang2024:PRB}.

\begin{figure}[t]
\centering
\includegraphics[scale=0.39]{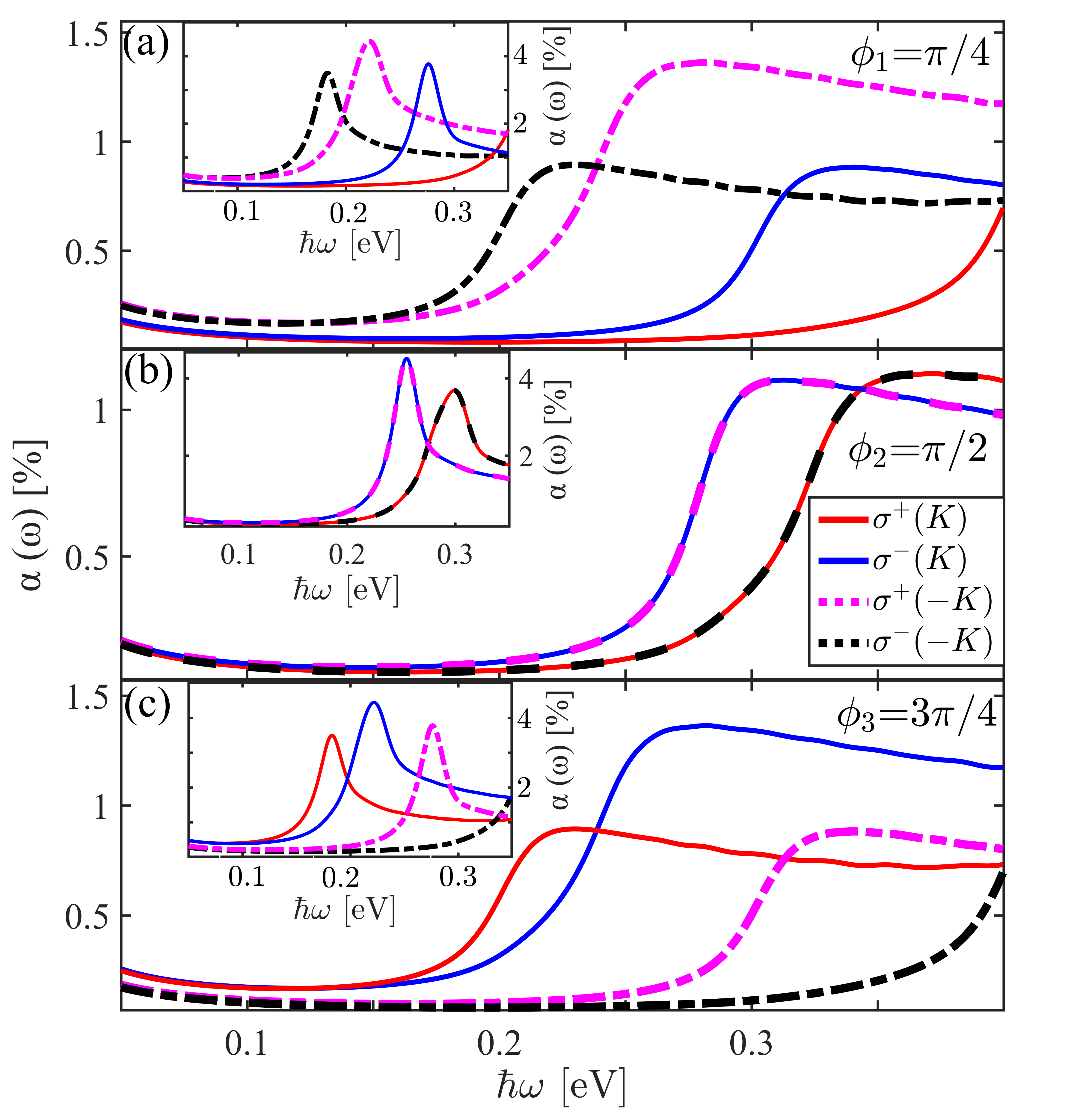}
\caption{ Absorption spectra of a ML SbH with magnetization directions (a) $\phi_1=\pi/4$, (b) $\phi_2=\pi/2$, and (c) $\phi_3=3\pi/4$. Inset: corresponding excitonic absorptions. }
\label{fig4}
\end{figure}

To achieve a more complete understanding, we investigate the  optical absorption associated with the photocurrent response. As shown in Fig.~\ref{fig3} (a), the CBs cross at $\phi_{c1}=1.1$ and $\phi_{c2}=2.0$, which divide $\phi$ into three intervals: (I) $0 \le \phi \le \phi_{c1}$, (II) $\phi_{c1}<\phi<\phi_{c2}$, and (III) $\phi_{c2} \le \phi \le \pi$. Correspondingly, we present in Fig.~\ref{fig4} absorption spectra at $\phi_1=\pi/4$, $\phi_2=\pi/2$, and $\phi_3=3\pi/4$, each representing one of the intervals. 
For $\phi_2=\pi/2$, i.e., in-plane field, the absorption at photon energy $\hbar \omega=0.29$ eV, which involves transition between higher VB and lower CB, exhibit a valley degeneracy and a valley-contrast circular dichroism: absorption at the $K$ ($-K$) valley is dominated by $\sigma^-$ ($\sigma^+$), as depicted in Fig.~\ref{fig1} (c). Note that while a simplified two-band model is applied for the photocurrent, the four-band Hamiltonian is considered in the absorption for completeness. Therefore, the absorption curve that emerge at the lowest energy corresponds to the two-band model and additional absorption curves appear at higher energies due to the involvement of higher CB and lower VB. 
For $\phi_1=\pi/4$, the absorption with the lowest energy appear at the $-K$ valley due to a valley splitting, with an opposite helicity due to the crossing of CBs at $\phi_{c1}$. Similarly, for $\phi_3=3\pi/4$, absorption emerges at the $K$ valley with a reversed helicity compared to that for $\phi_2=\pi/2$. Therefore, both photocurrents and absorption can be effectively tuned by the magnetization. Meanwhile, such changes of helicities are accompanied by sign changes of Berry curvature, which occur after each band crossing at the specific valley~\cite{Note:SM}. 

Since optical absorption in 2D materials are dominated by excitons, we further examine the excitonic absorption by numerically solving the BSE. Fig.~\ref{fig4} shows clear connections between the excitonic absorption peaks and the single-particle absorption. The helicities of the lowest excitonic absorption peaks, which are associated with the lower CB and higher VB, also get reversed after each band crossing. Notably, the energies of the excitonic peaks are lower compared to those of the respective single-particle absorption, due to finite exciton binding energies, which suggest lower photon energies in experiments. 





To conclude, we propose that magnetic effects can effectively break the symmetry in energy dispersions of centrosymmetric materials to generate helicity-dependent photocurrents, such that the limitation of inversion symmetry breaking in materials as a prerequisite of CPGE can be overcome. Generally, while such a MCPGE dominates the injection currents in centrosymmetric materials, it can also induce an extra externally tunable contribution to photocurrents in non-centrosymmetric materials subject to magnetic effects. Further tunability of the responses in both photocurrents and absorption can be achieved by controlling the magnitudes and directions of the magnetic exchange fields. Conversely, the response in photocurrents could also serve as a probe of the magnetic proximity effect in 2D materials. In addition to a defect-gradient approach, MCPGE can be employed to achieve helicity-dependent terahertz emission in centrosymmetric Dirac semimetal thin films~\cite{Chen2024:NC}. In addition to optical absorption, excitonic effects could be further explored to account for possible enhancement in photocurrent~\cite{Ruan2024:NL, Esteve2025:npj}. We expect that our scheme  could be verified experimentally with progresses in the synthesis of antimonene~\cite{Cheng2023:PCCP, Li2025:AM} and extended to other 2D materials including 2D altermagnets~\cite{Ma2021:NC, Jiang2025:NP, Zhu2025:NL} and topological insulators~\cite{Ma2021:NM}.


\begin{acknowledgments}

\end{acknowledgments}

\end{document}